\documentclass[10pt,conference]{IEEEtran}
\usepackage{graphicx}
\usepackage{xcolor}
\usepackage{bm}
\usepackage{booktabs}
\usepackage{amsmath}
\usepackage{amssymb}
\usepackage{makecell}
\usepackage{pifont}
\usepackage{mathtools}
\usepackage{subcaption}
\usepackage{xspace}
\usepackage{adjustbox}
\usepackage{comment}
\usepackage{stfloats} 
\usepackage{xurl}

\newcommand{\sys}{StreamMUSE\xspace}
\newcommand{\seqM}{\mathbf{M}}
\newcommand{\seqA}{\mathbf{A}}
\newcommand{\oldversion}[1]{}

\newcommand{\ziyunote}[1]{\textcolor{blue}{\textit{[\textbf{Ziyu's Note}] #1}}}

\usepackage{siunitx}
\IEEEoverridecommandlockouts

\begin{document}

\title{Real-Time Language Model Jamming:
A Case Study for Live Music Accompaniment Generation}

\author{
    \IEEEauthorblockN{
        Bowen Zheng$^{1,2,*, \ddagger}$,
        Andrew H. Yang$^{3,2,*, \ddagger}$,
        Jiaqi Ruan$^{4,2}$,
        Jia He$^{4,2}$,
        Xinyue Li$^{2}$,\\
        Yuan-Hsin Chen$^{5,2, \ddagger}$,
        Ziyu Wang$^{6,2,\dagger}$,
        Xiaosong Ma$^{2,\dagger}$
    }
    \IEEEauthorblockA{
        $^{1}$University of Wisconsin--Madison, USA, \texttt{bzheng68@wisc.edu}\\
        $^{2}$Mohamed bin Zayed University of Artificial Intelligence, UAE, \\
        \texttt{\{Jiaqi.Ruan, Xinyue.Li, Ziyu.Wang, Xiaosong.Ma\}@mbzuai.ac.ae}\\
        $^{3}$University of California, San Diego, USA, \texttt{any012@ucsd.edu}\\
        $^{4}$University of Science and Technology of China, China, \texttt{hej148@mail.ustc.edu.cn}\\
        $^{5}$Wuhan University, China, \texttt{yuanhsin@whu.edu.cn}\\
        $^{6}$New York University, USA
    }
    \thanks{$^*$Equal contribution. $^\dagger$Corresponding authors. \par$\ddagger$
    This project was supported by the MBZUAI UnderGraduate Research Internship Program (UGRIP) 2025, during which a substantial portion of the research was conducted.}
}

\maketitle

%
\begin{abstract}
Language models (LMs) have become one of the most prominent paradigms in modern generative modeling. While making them faster has been the main focus of real-time deployment, speed alone is not enough. Many real-world applications, such as synchronized translation and voice synthesis, also require precise alignment between generation and external signals, both in terms of generation content and timing. We refer to this problem as \textit{frame-synchronous streaming inference}. To address it, we present \sys, an inference system that performs LM generation in response to an external signal stream within a client-server architecture. The client continuously sends high-frequency inference requests based on the most recent inputs and receives outputs synchronized to the external clock, while the server executes model inference. We demonstrate the framework through a live music accompaniment task, showing how real-time synchronization can be achieved across different deployment environments with varying round-trip latencies. We further model the relationship between system hyperparameters and round-trip latency, and evaluate how different environments affect optimal configurations to achieve real-time performance. Experimental results show a consistent correspondence between system real-time performance and music quality, demonstrating the effectiveness of the proposed framework. The project is open source. Relevant code and the latest updates are available at \url{https://stream-muse-webpage.vercel.app/#audio-library}.
\end{abstract}







\section{Introduction}
\label{Sec:introduction}

Language models (LMs) have become the foundation of modern generative systems across modalities, from text to speech and music. As their applications expand into interactive and real-time domains, such as live conversation, synchronized translation, and music editing, the primary goal has often been to make inference faster ~\cite{fu2025efficientadaptivesimultaneousspeech, kasai2021deepencodershallowdecoder, ren2022fastspeech2fasthighquality}. Yet speed alone does not tell the full story. Consider the problem of improvising music with other musicians: the system must not only generate notes quickly, but also update its context frequently to ensure high-quality predictions and stay aligned with the shared musical clock. This highlights a broader challenge in real-time generative modeling---producing outputs that are \textit{synchronized in time} and \textit{coherent in content} with continuously evolving external signals.

In this work, we frame this challenge as \textit{frame-synchronous streaming inference}, where a language model $P_\theta(y|x)$ must generate an output sequence $y$ that evolves in real time with an external signal stream $x$. The generation proceeds at a fixed frame rate, synchronized to physical time, so that at each frame $t$, the model produces $y_t \sim P_\theta(y_t \mid x_{<t}, y_{<t})$ conditioned on the most recent external information $x_{<t}$ and its own generation history $y_{<t}$. 
Existing approaches, however, typically operate at much coarser or irregular temporal resolutions. They either assume long frames to predict low-resolution signals (e.g., 2-second chunks~\cite{dhariwal2020jukeboxgenerativemodelmusic}) or use dynamically triggered updates tied to user interactions (e.g., chatbots that respond after full sentences~\cite{thoppilan2022lamdalanguagemodelsdialog, shuster2022blenderbot3deployedconversational}). In contrast, frame-synchronous streaming inference targets short frame intervals (e.g., within 200 ms), comparable to the smallest perceptual or structural unit in music or speech (such as a tatum or syllable), where precise timing and synchronization are essential. In our music setting, this temporal unit is instantiated as a tick, while the corresponding symbolic representation at each tick is encoded as a frame.

In this paper, we present \sys, a system for frame-synchronous streaming inference. We take real-time music accompaniment generation as a case study, where the model generates accompaniment conditioned on a user-input melody that serves as the external signal. \sys operates in a client–server architecture: the client continuously sends high-frequency inference requests based on the latest user input, and the server performs language model inference and returns generation results aligned to the music stream. The system is characterized by two key parameters: the \textit{inference interval} ($I$), which determines how frequently requests are sent, and the \textit{generation length} ($GL$), which specifies how long each request generates output. These hyperparameters jointly balance responsiveness and reliability under stochastic round-trip latency, which depends on model inference time and network conditions between client and server.

To analyze this trade-off, we develop a latency-aware formulation that quantifies the feasible configurations of these parameters under varying latency conditions. A straightforward intuition is that real-time inference can be achieved by minimizing the inference interval and setting the generation length just long enough to cover each frame. In practice, however, random latency fluctuations make hyperparameter configuration non-trivial. For instance, an overly short $I$ can cause overlapping requests that stall the system, whereas a slightly longer $GL$ provides necessary temporal redundancy as a backup against delayed requests. We further examine how an increment of music tempo---the innate music concept that defines the frame rate---tightens these constraints, sometimes even making real-time scheduling infeasible.

We train a real-time accompaniment generation model as the backbone LM and evaluate \sys across three environments---local, local-server, and remote-server---with varying network latencies and inference speeds.
Experimental results show that system behavior aligns well with our model for hyperparameter selection, and that the system responsiveness is strongly correlated with music-quality metrics. The experiments further confirm that, when feasible under given latency conditions,  more frequent inference requests and suitable generation length yield performance closest to the offline baseline, demonstrating the importance and necessity of high-resolution streaming inference.

\section{Background and Related Work}
\label{Sec:background}

This section reviews prior work on LLM inference, music generation, and real-time systems, with a focus on their applicability to the frame-synchronous streaming setting considered in this work.

\subsection{Language Model Inference}
Significant research efforts have been dedicated to optimizing and accelerating large language model (LLM) inference, primarily following two complementary technical pathways. The first focuses on architectural innovations, such as transforming dense models into sparse Mixture-of-Experts (MoE~\cite{shazeer2017outrageously}) structures, which preserve total parameter count while substantially reducing computational load per input token. The second approach retains the algorithmic foundation while optimizing resource utilization at the framework level—exemplified by systems like vLLM~\cite{kwon2023efficientmemorymanagementlarge} and SGLang~\cite{zheng2024sglangefficientexecutionstructured}, which introduce advanced techniques such as continuous batching, paged KV cache management, and adaptive scheduling. In hard real-time task scenarios such as automotive intelligent software~\cite{Vehicles}, systems must further incorporate task prioritization and resource management on top of model-level optimizations, ensuring critical functions (e.g., braking decisions) receive prioritized inference resources. In summary, existing optimizations are primarily designed for general-purpose language models, whose token-sequential generation process and heavy reliance on statistical dependencies make them difficult to directly apply in edge scenarios with strict requirements for real-time performance, determinism, and task criticality.

\subsection{Machine Learning for Music Generation}
Music has long been intertwined with real-time interaction, dating back to early research on score following~\cite{dannenberg1984realtime, orio2003scorefollowing}, timbre recognition~\cite{hung2018framelevel, esling2018generativetimbrespace}, and other interactive music systems that require instantaneous computational responses~\cite{rowe1993interactivemusic, pachet2003continuator, cont2010realtimemusic}. Accompaniment generation, a representative task in music AI, is particularly well suited for real-time musical interaction due to its inherently responsive and continuous nature.

Deep learning has significantly advanced symbolic music generation, with models such as Music Transformer~\cite{huang2019musictransformer} and Multitrack Music Transformer~\cite{dong2023mmt} excelling at modeling long-term structure and multi-instrument arrangements. Several studies have further explored controllable symbolic generation, such as melody-conditioned accompaniment. AccoMontage~\cite{zhao2021accomontageaccompanimentarrangementphrase} retrieves and adapts stylistic phrase representations from a precompiled corpus; Whole-Song Hierarchical Generation~\cite{wang2024wholesonghierarchicalgenerationsymbolic} generates accompaniment hierarchically from global structure to detailed parts; and the Anticipatory Music Transformer~\cite{thickstun2024anticipatorymusictransformer} improves quality through a look-ahead design that explicitly violates real-time assumptions. These methods produce musically coherent and controllable results, yet real-time generation remains challenging because they rely on offline processing and full-sequence context.
 
 \subsection{Real-Time Accompaniment Generation System}

 Despite major advances in deep learning for symbolic music modeling, its integration into real-time interactive systems has been relatively limited. Only a few recent studies have explored this direction, yet their focus largely lies on algorithmic design—particularly in reinforcement learning—rather than on the system-level challenges of achieving stable, low-latency interaction. 
 
  For instance, the RL-Duet framework~\cite{jiang2020rlduetonlinemusicaccompaniment} formulates real-time accompaniment as a sequential decision-making process using reinforcement learning, enabling the agent to respond instantaneously to a performer’s input; however, its discrete action formulation leads to coarse timing and limited expressiveness. Similarly, Bach-Duet~\cite{Benetatos2020BachDuetAD} uses an RNN for real-time counterpoint improvisation, but its strict 16th-note quantization and monophonic constraints lack the expressiveness required for rich, polyphonic accompaniment. More recently, ReaLJam~\cite{scarlatos2025} introduces a reinforcement-learning-tuned Transformer for interactive jamming with predictive scheduling, yet its design remains centered on chord- or measure-level anticipation. Meanwhile, the Jam\_Bot~\cite{blanchard2025jambot} enables human–AI free improvisation by adapting symbolic music language models; however, its architecture is highly artist-specific. It relies on fine-tuned stylistic data and handcrafted interaction strategies, which limits its generalization to broader real-time or multi-user scenarios.
 
 Even in recent industry-scale developments, such as Google Magenta's real-time live music models~\cite{caillon2025live}, the focus remains distinct. Their system targets a distinct set of objectives, focusing on adaptive music generation steered by high-level text or audio prompts, rather than the tight, frame-synchronous melody conditioning explored in this work. Furthermore, their models operate on two-second chunks, meaning user inputs take two or more seconds to influence the musical output. While such a design is effective for its intended use case, the resulting multi-second latency presents a hurdle for highly interactive scenarios that necessitate sub-200ms responsiveness.
 
 Overall, these works represent valuable early steps but remain coarse-grained in temporal resolution and system integration. In contrast, our work focuses on the tackling the system problem—building a responsive, low-latency, and robust real-time generation framework—addressing the fundamental gap between algorithmic sophistication and deployable real-time musical interaction.

\section{Model Overview}
\label{Sec:modelOverview}

\subsection{Statistical Model for Real-Time Accompaniment Generation}
\label{sec:statistical-model}

In this study, we formulate the real-time interactive accompaniment task as a \emph{conditional sequence modeling problem}. 
The core design principle, crucial for low-latency collaboration, 
is the online constraint: upon each generation request, the model must only generate new accompaniment 
tokens conditioned on the observed melody history and all previously generated accompaniment. 
This process is strictly causal and permits no lookahead into the user-fed future melody. In the generic streaming notation introduced earlier, $x$ corresponds to the melody sequence $\seqM{}$ and $y$ to the accompaniment sequence $\seqA{}$; correspondingly, $x_{<t}$ and $y_{<t}$ map to the histories $\seqM{}[0:t-1]$ and $\seqA{}[0:t-1]$ in Equation~\ref{eq:causal-model}.

Considering the incremental and rhythmic nature of real-time music generation, in this work we discretize time into uniform temporal steps called \emph{ticks}, as the finest resolution considered. 
Its size is configurable, though in our experiments a tick is fixed at 1/4 beat (1/16 note), its wall clock time duration $\tau_{\text{tick}}$ depends on the music speed (commonly measured in BPM, or Beats Per Minute). 
E.g., for a song of 120 BPM, $\tau_{\text{tick}}$ would be 125ms.

Formally, let $\seqM{}[i:j]$ denote the melody token sequence and 
$\seqA{}[i:j]$ denote the accompaniment token sequence starting at tick $i$ and ending at tick $j$, the model estimates the conditional 
distribution
\begin{equation}
    p(\seqA{}\mid\seqM{})\;=\;\prod_{t=0}^{T-1} p\big(\seqA{}[t] \mid \seqA{}[0:t-1],\, \seqM{}[0:t-1]\big).
    \label{eq:causal-model}
\end{equation}

In other words, for tick t, the model should predict $\seqA{}[t]$ prior to observing this segment melody $\seqM{}[t]$, so that $\seqA{}[t]$ can be produced in time to accompany it. 
A tokenization method that supports the real-time setting and model architecture is introduced later in this section.

\subsection{Data Representation}
\label{data_representation}

The data representation described above is a conceptual abstraction, to support real-time generation, we employ a more detailed and realistic music data encoding scheme. 
Based on the tick-based granularity introduced earlier, we represent the melody and accompaniment as two aligned sequences of \emph{frames}. Here, a frame denotes the tokenized representation at a single tick; throughout the rest of the paper, we use \emph{tick} to refer to the temporal/scheduling unit and \emph{frame} to refer to its symbolic encoding:
\begin{align*}
    \seqM{} &= (\seqM{}[0], \dots, \seqM{}[t-1]), \\
    \seqA{} &= (\seqA{}[0], \dots, \seqA{}[t-1]).
\end{align*}

To enable efficient autoregressive prediction, these two sequences are combined using an 
\emph{interleaving} technique into a single input sequence
$$
\mathrm{InterL}(\seqM{}, \seqA{}) = (\seqA{}[0], \seqM{}[0], \dots, \seqA{}[t-1], \seqM{}[t-1]).
$$

\begin{figure}
    \centering
    \includegraphics[width=0.99\linewidth]{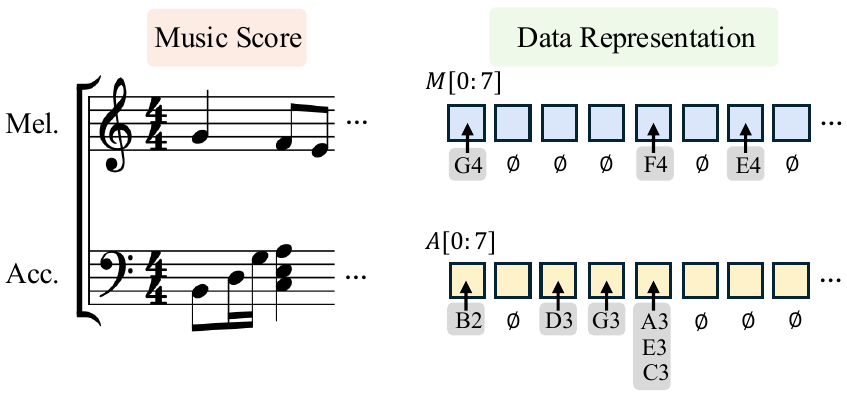}
    \caption{Sample MIDI tokenization}
    \label{fig:tokenization}
\end{figure}

Given the polyphonic nature of music, both a melody frame ($\seqM{}[t]$) and an accompaniment frame ($\seqA{}[t]$) denote the symbolic encoding of all note events occurring at tick $t$.
This allows a single frame to summarize multiple simultaneous notes (or silence). For example, an accompaniment frame can represent three simultaneous notes forming a chord. 
Fig.~\ref{fig:tokenization} illustrates such tokenization.
Here, the 2 beats of sheet music shown on the left side translate into 8 ticks on the right, where internal representations like ``B2'' define the pitch of notes.  
Each blue/yellow block in the figure represents a melody/accompaniment frame. 

Note that based on the polyphony encountered in our training and testing music datasets,
we empirically set the maximum number of note events per frame to 4, which is easy to extend if necessary.

\subsection{Model Architecture and Training}
\label{model_architicture}

Our model itself is a Transformer-based, autoregressive decoder conditioned on the interleaved melody and accompaniment
sequence. 
Its backbone architecture is adapted from the three-module design introduced in~\cite{jiang2025versatilesymbolicmusicformusicmodeling}, consisting of a local encoder, a global predictor, and a local decoder for efficiently processing structured symbolic music representations. 
The local encoder encodes notes starting at the same tick into a single token. The global predictor then processes these aggregated vectors across the entire history to capture global musical 
and rhythmic context and predict high level representation of the next token. Finally, the local decoder generates the new frame's subsequence autoregressively, 
conditioned on the global predictor's output for the current time step. 

The loss function used for training this model is the standard cross-entropy loss based on Equation \ref{eq:causal-model}. During training, the training data sequences start with the accompaniment token $\seqA{}[0]$. Since our model is trained with teacher forcing, the resulting discrepancy between training and inference—known as exposure bias—is an inherent challenge in this streaming setting.

\subsection{\sys Architecture Overview}
\begin{figure*}[t]
    \centering 

    \begin{minipage}{0.90\textwidth} 
        \centering 

    \centering
    \begin{subfigure}[b]{0.45\textwidth}
        \centering
        \adjincludegraphics[
            trim={0 0 {0.52\width} 0},
            clip,
            height=8cm
        ]{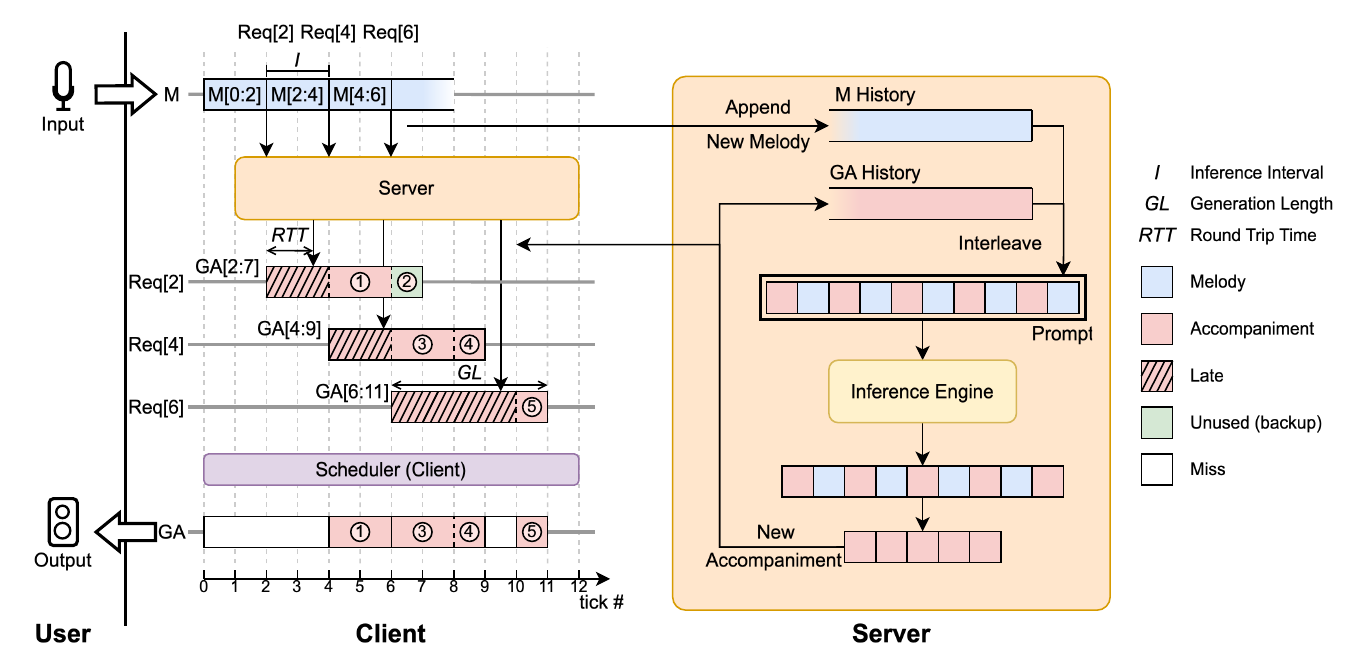}
        \caption{Request timeline}
        \label{fig:diagramA}
    \end{subfigure}
    \hfill
    \begin{subfigure}[b]{0.54\textwidth}
        \centering
        \adjincludegraphics[
            trim={{0.48\width} 0 0 0},
            clip,
            height=8cm
        ]{fig_topdown_overview.pdf}
        \caption{Server workflow}
        \label{fig:diagramB}
    \end{subfigure}
    \caption{\sys architecture. The left panel shows a sample timeline with three requests: Request-1 (Req[2]), Request-2 (Req[4]), and Request-3 (Req[6]), issued at the onset of ticks 2, 4, and 6, respectively. The right panel gives a zoomed-in view of the server-side workflow corresponding to one request.}
    \label{fig:bothDiagrams}

    \end{minipage}
\end{figure*}

Finally, we assemble the above building blocks and give a sketch of the architecture of our proposed \sys system. 
While the above model allows incremental accompaniment generation,  achieving real-time accompaniment is far from straightforward. 
Live accompaniment is highly sensitive to latency and BPM (A higher BPM accelerates the musical rhythm, leading to more rhythmically discordant beats under identical latency conditions.) 
Only when the accompaniment closely follows the performer’s melody does the overall performance sound natural. Yet, the round trip time ($RTT$) of fulfilling each request due to network latency and server side inference is significant, making it difficult for the system to produce perfectly synchronized accompaniment in real time.

To enable \sys's working with diverse hardware and service settings, we adopt the common client-server design, as illustrated by Fig.~\ref{fig:bothDiagrams}.
Such a distributed architecture also decouples latency-sensitive user interaction from computation-intensive model inference.

The client, which takes user input MIDI melody, sends accompaniment generation requests at a certain frequency, again defined by tick intervals ($I$ ticks). 
The server processes a request, generating an interleaved sequence (using the $\mathrm{InterL}$ function), whose size is specified by the generation length parameter $GL$. 
It is desirable to have a $GL$ long enough to produce accompaniment sequences overlapping in time, so that there are backup segments in case of network or inference latency jitters. 
The accompaniment component of the generated sequence, GA, is then returned to the client, which coordinates the responses from multiple outstanding requests and schedules the accompaniment playback. 
Both the ground-truth (user-fed) melody and the GA results are maintained by the server side for subsequent inferences. 

\begin{table}[h!]
\centering
\setlength{\tabcolsep}{2pt} 
\caption{Key Definitions for \sys}
\renewcommand{\arraystretch}{1.2}
\footnotesize
\begin{tabular}{p{2.5cm}p{5cm}}
\toprule
\textbf{Symbol} & \textbf{Description} \\
\midrule
$\seqM{}, \seqA{}$ & Sequences of melody and accompaniment frames, one per tick \\
$\mathrm{BPM}$ & Beats per minute \\
$\mathrm{tick}$ & Minimum work unit in \sys's processing, set as a fraction of a beat\\
$\tau_{\text{tick}}$ & Tick duration in milliseconds \\
$\mathrm{InterL}(\seqM{}, \seqA{})$ & Interleaved melody–accompaniment token sequence \\
$\mathcal{L}$ & Cross-entropy loss function \\
$I$ & Inference interval \\
$GL$ & Generation length \\
$\mathrm{GA}$ & Generated accompaniment \\
$\mathrm{Req}$ & Request \\
$\mathrm{RT}_{\text{tick}}$ & Round-trip time in tick unit \\
$RTT$ & Round-trip time in milliseconds \\
\bottomrule
\end{tabular}
\label{tab:defs}
\end{table}
Table~\ref{tab:defs} summarizes the key definitions introduced, to be used throughout the paper.
In Section~\ref{Sec:StreamMUSE}, we revisit Fig.~\ref{fig:bothDiagrams} and give detailed description of the design of our client and server. In particular, how they work together to automatically search for optimized $I$ and $GL$ parameter values under a given hardware/service setup.


\section{\sys Design and System Modeling}
\label{Sec:StreamMUSE}

\subsection{Client Design}
The client employs a multi-threaded design and an asynchronous scheduling mechanism that decouples real-time playback from network communication. The Main Thread coordinates timing and playback scheduling, the Input Thread captures user melody input in real time, and the Inference Thread manages communication with the server. At each time interval (typically consisting of several ticks), the client packages the latest melody segment and sends it to the server for inference. It then receives a generated accompaniment of length $GL$ from the server for playback scheduling.

In practical network modeling, we have observed that network latency follows a long tail distribution, meaning a small fraction of requests exhibit round-trip times ($RTT$) significantly higher than the average. To address this, we implemented a tick-aligned data structure and introduced a backup mechanism as a safety buffer. The generated accompaniment (GA) returned by each server is stored in this data structure for real-time scheduling. By setting the generation length ($GL$) to be longer than the inference interval ($I$), an overlap is created between consecutive GAs. This overlapping portion serves as a backup left by the previous request and will be overwritten by data from a new, on time request. If a new request is delayed, the backup is scheduled in real-time to maintain musical continuity. A missing accompaniment event occurs only when a request is delayed and no backup accompaniment is available in the system.

In the sample scenario shown in Fig. \ref{fig:bothDiagrams}(a), $I$ is set to 2 ticks and $GL$ is set to 5 ticks (corresponding to 9 interleaved frames). With an experienced $RTT$ of 1.5 ticks, the Scheduler receives the return value of Request-1 and writes it to GA[2:7]. Due to the upward rounding of ticks (where a tick is the smallest scheduling unit of the accompaniment), the shaded portion of GA in Fig.~\ref{fig:bothDiagrams}(a) is already too late to be scheduled. GA\ding{172} can be directly scheduled in real-time, while GA\ding{173} serves as a backup for subsequent accompaniment. Since Request-2 returns as expected, the generated accompaniment is overwritten, and the newly returned GA\ding{174} will be scheduled, with GA\ding{175} acting as the backup. If Request-3 is delayed, the backup from the last request GA\ding{175} will be scheduled. However, if the delay exceeds its coverage, no available GA remains for scheduling, resulting in a miss until Request-3 returns.

The backup mechanism effectively mitigates the impact of network tail latency. 
However, having excessively long backup generation is clearly not advantageous. Although it can reduce the risk of missing accompaniment, it also increases the inference latency for all requests, potentially compromising the system’s real-time constraints.
Hence, the goal of our real-time system is twofold: (1) to ensure musical continuity by minimizing the number of missing events (the white ``hole'' at tick 9 in Fig.\ref{fig:bothDiagrams}), and (2) to satisfy the real-time constraint by reducing reliance on backup accompaniment. 
Later in this section, we further formalize the 
real-time behavior of \sys. 

\subsection{Server Design}
\label{sec:4c}
The core of the server is an inference engine built on the generative model described in Section~\ref{Sec:modelOverview}. As shown in Fig.~\ref{fig:bothDiagrams} (b), during online generation, the server maintains a tick-aligned data structure to store the user's historical generated accompaniment and real melodic context information. Upon receiving the real-time melody segment from the client, the server appends it to the existing historical melody data. For instance, as shown in Fig. 2, Request-2 only transmits the incremental segment $\seqM$[2:4], which the server then integrates into its melody history for subsequent inference. The accompaniment and melody at the same tick are then interleaved to form a new sliding window, which is fed into the model as a unified melody–accompaniment sequence prompt. The model infers and generates the next segment of the melody–accompaniment (9 frames in this example), where the 5 frames/ticks of accompaniment result are returned to the client, as well as saved as accompaniment history. 

After inference, the server separates the generated melody and accompaniment, discards the generated melody, retains the generated accompaniment, and transmits it to the client. This mechanism minimizes data transmission between client and server while maintaining real-time updates of the user's melody, ensuring both real-time performance and the accuracy and contextual relevance of the generated results.

\subsection{Round-Trip Time Modeling}
As previously discussed, $RTT$ is an inherent system latency that fundamentally determines the theoretical upper bound of real-time performance in online music generation. To model $RTT$, our analysis decomposes it into two core components: (i) inference latency ($IL$), which refers to the computational time required on the server side to complete the music generation; and (ii) network latency ($NL$), which encompasses the communication overhead incurred during end to end transmission of the request and response data.

To model $IL$, the key relevant factor is the model’s autoregressive generation length ($GL$), as  
a request's $RTT$ can be formally expressed as
$$
RTT(GL) = IL(GL) + NL.
$$
 
Across multiple hardware and inference service setups, we sweep the typical 
values $GL$ and record $IL$ to analyze their relationship. Through our detailed breakdown, 
we find that in the current three module structure, $IL$ is dominated by the local decoder that 
autoregressively generates frames. Within the computational complexity of the local decoder, the 
leading term is determined by the attention mechanism, whose cost scales quadratically with $GL$. 
Accordingly, we model $IL$ as a quadratic function of $GL$ with an additive Gaussian noise term:
\[
IL = f(GL) + \epsilon, \quad \epsilon \sim \mathcal{N}(0, \sigma^2),
\]
where $f(GL) = \alpha GL^2 + \beta GL + \gamma$ is a quadratic function of $GL$, with $\gamma$ denoting the constant term, and $\epsilon$ represents stochastic fluctuations in inference latency caused by system load and other runtime uncertainties.


To model $NL$, in a local setup where the client and server run on the same host, there is no network communication, so we set $NL$ to $0$; with the other network setups, $NL$ fluctuates around a constant value due to fixed communication overhead and unstable wireless links. To capture the heavy tailed behavior of network latency, we replace the Gaussian noise with a shifted power law (Pareto) noise term:
\[
NL =\nu + (\xi - x_{\min}), \quad \xi \sim \mathrm{Pareto}(x_{\min}, \alpha),
\]
where $\nu$ is a constant representing the fixed communication delay, $x_{\min}>0$ is the scale (minimum) parameter, and $\alpha>1$ is the tail index controlling the heaviness of the tail. The probability density function (PDF) of $\xi$ is given by
\[
g_{\xi}(x) = \frac{\alpha\,x_{\min}^{\alpha}}{x^{\alpha+1}}, \quad x \ge x_{\min}.
\]
This formulation ensures that the noise term $(\xi - x_{\min})$ has a minimum value of zero, allowing $NL$ to fluctuate around $\nu$ while exhibiting occasional extreme latency spikes consistent with heavy tailed network behavior. We emphasize that the Pareto assumption is used as a practical approximation of upper-tail latency behavior rather than a claim of exactness under all network conditions. Empirically, the tail regime of our measured $RTT$ traces is well approximated by a linear fit on log-log plots, with an average coefficient of determination $R^2 = 0.949$ across the evaluated settings. We therefore adopt Pareto as a conservative engineering choice, since it implies larger buffer requirements than lighter-tailed alternatives and thus provides a safety margin for real-time parameter selection.

Putting the above together, we write $RTT$ in explicit form as
$$
RTT = f(GL) + \nu + (\xi - x_{\min}) + \epsilon.
$$


\subsection{Model-Based \sys Parameter Configuration}
\label{sec:param_selection}

Finally, we come to the key problem of a real-time interactive music accompaniment system, to balance generative quality, interactive responsiveness, and the physical constraints of computation and networking. 
To this end, our request response time modeling given above allows us to move beyond ad-hoc parameter tuning, adopting instead a formal methodology that derives the feasible solution space of real-time safe configurations. 
In this manner, under different hardware, network, and inference service configurations, \sys automatically identifies the optimal operating point to maximize interactivity and reliability.

The derivation depends on relating the system's temporal latency (in milliseconds) to the musical ``tick'' (our discrete time unit). We first introduce a key parameter:
\begin{itemize}
    \item $\tau_{tick}$ (Time per Tick): The duration of a single tick in seconds. A 16th note resolution (4 ticks per beat), the time per tick is derived from BPM:
    $$\tau_{tick}=\frac{60 \times 1000}{4 \cdot BPM}   (\text{ms/tick}).$$
\end{itemize}
For this constraint analysis, we simplify the RTT model from Section~4.2 by absorbing the constant term $\gamma$ in $f(GL)$ together with $\nu$, $(\xi - x_{\min})$, and $\epsilon$ into a single constant term $c$ for tractability. This yields the following approximation for the 95th-percentile tick-level response time:
$$Q_{0.95}(\{\gamma, \nu, \xi - x_{\min}, \epsilon\}) \le c,$$
\begin{equation}
\label{eq:rt_tick_95}
RT_{tick}^{95} \approx \frac{f(GL) + c}{\tau_{tick}}.
\end{equation}

With these variables defined, we establish two fundamental constraints for real-time operation:
\begin{enumerate}
    \item Real-Time Safety Constraint: The client should ideally receive the server generated accompaniment before the next interval begins:
    \begin{equation}
    \label{eq:rt_safety}
    \lceil RT_{tick}^{95} \rceil \le I.
    \end{equation}
    \item Playback Continuity Constraint: The duration of the generated accompaniment must be designed to accommodate outlier $RTT$ instances present in the long tail segment of the latency distribution:
    \begin{equation}
    \label{eq:playback_continuity}
    GL \ge I + \lceil RT_{tick}^{95} \rceil.
    \end{equation}
    Rearranging this inequality to solve for $I$, we get the form used in our final derivation:
    \begin{equation}
    \label{eq:playback_continuity_rearranged}
    I \le GL - \lceil RT_{tick}^{95} \rceil.
    \end{equation}
\end{enumerate}


By combining our two constraints, we derive a single, two-sided inequality that \textit{defines the valid range for} the \textit{Inference Interval ($I$)} and substituting our RTT model $f(GL)$ into it gives the fully expanded constraint:
\begin{equation}
\label{eq:constraint_I_expanded}
\begin{split}
    \frac{f(GL) + c}{\tau_{tick}} 
    &\le I \le \frac{\tau_{tick}GL-f(GL) - c}{\tau_{tick}}.
\end{split}
\end{equation}

In our experimental environment, the behavior of $f(GL)$ can be accurately modeled using a quadratic function. For any given network condition, we first determine the simulated RTT function through benchmarking. Then, for a specified BPM (which defines the value of $\tau_{tick}$), we iterate over a range of possible Generation Lengths ($GL$) and compute the lower and upper bounds from Equation~\ref{eq:constraint_I_expanded}. If the lower bound is less than or equal to the upper bound, the resulting integer range of $I$ defines valid ($I$, $GL$) configuration pairs. Repeating this process over candidate $GL$ values delineates the feasible solution space, which can then be evaluated under additional criteria such as choosing the smallest $I$ for responsiveness or selecting specific configurations to reduce server load.


As shown later in Fig.~\ref{fig:solution_space}, the feasible solution space can vary substantially across deployment environments. In our setting, however, feasibility alone is not enough: although larger $I$ and $GL$ values can make the system easier to schedule, they also reduce interaction frequency and degrade response freshness, which is musically undesirable in live accompaniment. Therefore, valid configuration pairs must be evaluated together with the music-quality results in Table~\ref{tab:quality-results}. In practice, our goal is to keep $I$ and $GL$ as small as the system allows while preserving responsiveness, freshness, and musical coherence.

A related but simpler question is whether real-time inference is possible at all for a given model and deployment environment. From \eqref{eq:rt_safety} and \eqref{eq:playback_continuity_rearranged}, a feasible integer $I$ can exist only if the lower bound does not exceed the upper bound. In other words, the minimum admissible $I$ from the Real-Time Safety Constraint must not exceed the maximum admissible $I$ from the Playback Continuity Constraint. This yields the following necessary condition for real-time feasibility.

\textbf{Corollary (Feasibility Condition).}
A necessary condition for system feasibility is
\[
GL \ge 2 \left\lceil RT_{\text{tick}}^{95} \right\rceil.
\]
Using the approximation in \eqref{eq:rt_tick_95}, this condition can be written approximately in milliseconds as
\[
RTT(GL) \lesssim \frac{\tau_{\text{tick}}}{2} GL.
\]
This means that real-time feasibility requires the $RTT(GL)$ curve to remain below a linear upper bound with slope $\frac{\tau_{\text{tick}}}{2}$; since $\tau_{\text{tick}}$ decreases as BPM increases, this bound becomes tighter at faster musical tempos.
\section{Evaluation}
\label{Sec:evaluation}

\begin{table}[t]
  \centering
  \caption{Specifications of three target settings}
  \label{tab:settings}
  \resizebox{\linewidth}{!}{
  \begin{tabular}{@{}cccc@{}} %
    \toprule
    Setting & Client & Server & Network \\
    \midrule
    Local & \makecell{Hyperstack server\\(RTX A4000)} & \makecell{Hyperstack server\\(RTX A4000)} & IPC \\
    Local-server & \makecell{Macbook\\(M1 Pro)}   & \makecell{PC\\(RTX 3090)} & 6Gbps WLAN \\
    Remote-server & \makecell{Macbook\\(M1 Pro)}   & \makecell{Hyperstack server\\(A100)} & 1Gbps WAN \\
    \bottomrule
  \end{tabular}
  }
\end{table} \subsection{Experimental Setup and Model Preparation} 
\textbf{Implementation} We build our sample real-time music accompaniment system via Python with more than 5000 lines of codes, containing the full implementation of client-server architecture. For client side, we use multi-threaded mechanism to implement its different component; for server side, we select Transformers~\cite{wolf-etal-2020-transformers} library from HuggingFace as the inference engine, and enable KVCache optimization~\cite{pope2023efficiently} to avoid redundant computation of the autoregressive generation of local decoder. Client and server use TCP protocol to communicate with each other when network is needed.\\
\textbf{Testbed} We analyze the performance of our system under three different settings:
\begin{itemize}
  \item \textbf{Local} Client and server reside on the same device, directly communicating with each other via inter-process communication (IPC).
  \item \textbf{Local-server} Client and server are hosted on separate devices within the same wireless local area network (WLAN).
  \item \textbf{Remote-server} Client resides in a home network, while the server is deployed on a public cloud platform (e.g., Hyperstack). Communication between them is established via a wireless wide area network (WAN).
\end{itemize}
For the listed three settings, we summarize the hardware and network specifications in Table \ref{tab:settings}.\\
\textbf{Baselines} For the real-time performance, since we are the first to rigorously model real-time interactive transformer system, we mainly compare the performance of different \sys's configs within the search space to demonstrate the effectiveness of our parameter selection. We also compare the music quality of \sys\,with generated music from the offline version of our accompaniment generation model, which is used as the theoretical upper bound for interactive music quality.\\
\textbf{Training Details} We use the POP909 dataset \cite{pop909-ismir2020} for training, consisting of 909 pop music pieces. The model is trained using the standard cross-entropy loss on the interleaved token sequence. We also apply pitch shifting for data augmentation, where the shift amount is uniformly sampled from the allowed range (clipped to $\mathbf{[-5, 6]}$ semitones) for both melody and accompaniment tracks, and use gradient clipping to stabilize training. For evaluation, we primarily utilize the Small model (0.12B parameters). Its three-module architecture consists of a 12-layer main module alongside a 3-layer local encoder and a 3-layer local decoder.\\
\textbf{Evaluation Song Selection} Our conditional sequence modeling task inherently requires music data where melody (mel) and accompaniment (acc) are cleanly separated into distinct tracks. However, most publicly available MIDI datasets do not provide this essential separation.
To address this common challenge, we adopt the standard industry practice of skyline extraction. This method heuristically extracts the most prominent melodic line from a polyphonic music track, ensuring the necessary separation for our model's conditioning.
For our experiments, we constructed a dedicated test set with 64 music pieces by sampling 30 from AccoMontage and 34 from the POP1K7 dataset~\cite{pop1k7:compoundword2021}.\\

\begin{figure}[htbp]
    \centering
    \includegraphics[width=0.5\textwidth]{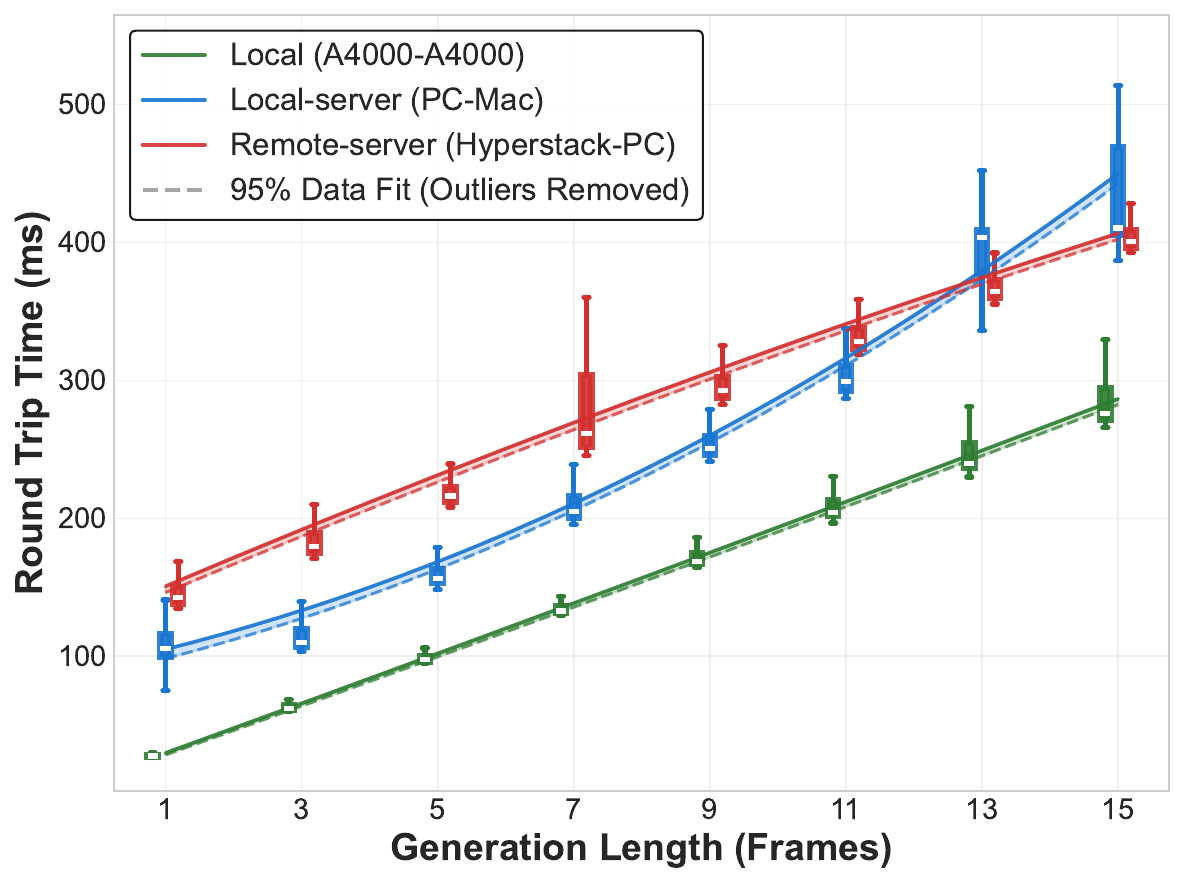} 
    \caption{Comparison between RTT model and real $RTT$ latency}
    \label{fig:rt_valid}
\end{figure}

\subsection{$RTT$ Latency Characteristics}
\label{sec:5b}

To understand the $RTT$ patterns of the \textit{Local}, \textit{Local-server}, and \textit{Remote-server} environments, we profiled them by varying the generation length ($GL$) and collecting 1,000 $RTT$ samples for each. The resulting distributions are shown in Fig. \ref{fig:rt_valid}.

Rather than simply validating our model, we use our quadratic RTT model as a tool to characterize and quantify each environment's performance. As shown in Fig. \ref{fig:rt_valid}, we fit parameters ($\alpha, \beta, \gamma$) to both 100\% (solid lines) and 95th percentile (dashed lines) of the data. Here, $\alpha$, $\beta$, and $\gamma$ denote the quadratic, linear, and constant coefficients of the fitted RTT function, respectively. We use the 95th-percentile fit for our analysis, as it avoids skew from rare outliers and provides the robust $RT_{\text{tick}}^{95}$ boundary essential for our safety constraints (Section~\ref{sec:param_selection}). This analysis reveals three distinct profiles:

\begin{itemize}
    \item Local (Green) serves as our computational baseline, isolating performance from network variability. It exhibits a near-zero base latency ($\gamma$) and minimal variance, with $RTT$ dominated by predictable computational cost.

    \item Local-server (Blue) has a lower base latency ($\gamma \approx 100 \text{ms}$) than Remote, but its $RTT$ scales more rapidly, overtaking the Remote Server at $GL \approx 13$. This environment is also highly unstable, with $RTT$ variance increasing dramatically with $GL$.
    
    \item Remote-server (Red) has the highest base latency ($\gamma \approx 150 \text{ms}$) due to network distance, but its $RTT$ scales slowly, confirming its powerful GPU. This environment is high-latency but predictable, with stable variance (barring $GL=7$).
\end{itemize}

\subsection{Visualization of Solution Space}
\begin{figure*}[t]
    \centering
    \includegraphics[width=\textwidth]{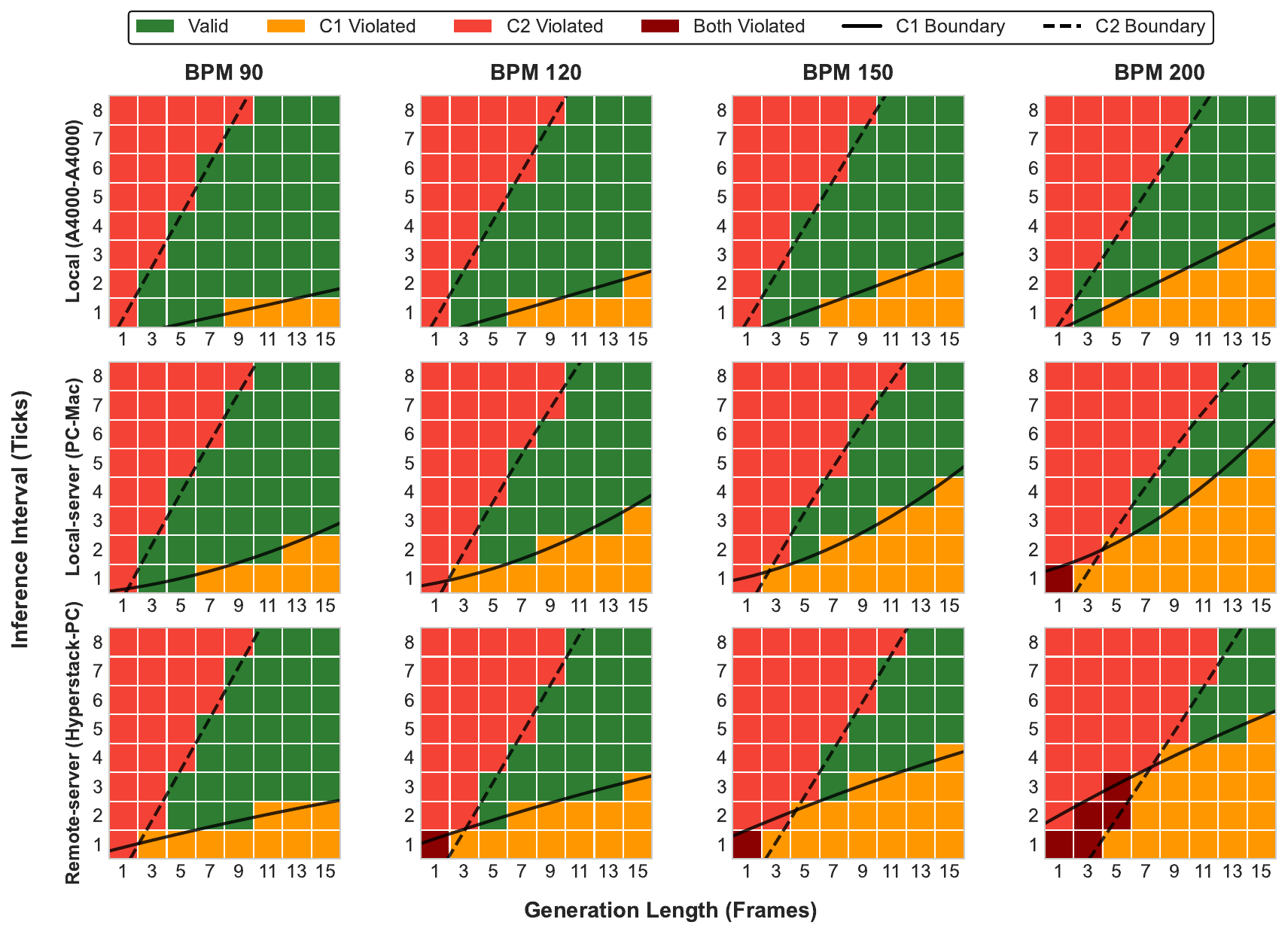} 
    \caption{Visualization of possible configuration pairs in solution space across three settings and four BPMs (Green: valid pair; Yellow: pairs violate C1 constraint; Red: pairs violate C2 constraint; Dark Red: pairs violate both constraint)} 
    \label{fig:solution_space}
\end{figure*}

After validating the fitting of our RTT model, we use RTT model with the constraints described in section~\ref{sec:param_selection} to search for all possible ($I, GL$) configuration pairs that fulfill real-time requirements under three settings, which is defined as the solution space of \sys. After getting the solution space, we plot all the feasible real-time configuration pairs explored by our system under Local, Local-server, and Remote-server settings across different BPMs, as illustrated in Fig. \ref{fig:solution_space}. The search space is defined by generation length ($GL$) and inference interval ($I$). \sys constructs the feasible solution space bounded by a solid line (representing the Real-Time Safety Constraint) and a dashed line (representing the Playback Continuity Constraint).

In different settings, variation in hardware and network conditions for interactive system may lead to situations where real-time requirements are violated. For instance, when the $RTT$ exceeds the interval $I$, the client may issue requests at a faster rate than the server’s response capacity, resulting in a mismatch between the returned accompaniment and the user’s current playing melody—referred to as C1 violation. Alternatively, if $GL$ is shorter than the sum of $RTT$ and $I$, user may experience missing intervals of accompaniment during playing—referred to as C2 violation. It is clear from Fig. \ref{fig:solution_space} that \sys can consistently find solution space in different settings, and the solution space varies across settings. In local setting, since the network latency is completely avoided, solution space can occupy no less than  44\% of the total search space. On the contrary, in remote-server setting, over 60\% of configurations within the search space exhibit violations of real-time conditions, making it challenging to manually construct a configuration that satisfies real-time requirements. Moreover, the increasing of the BPM exacerbates the violation, shrinking the solution space to 71\%, 33\%, and 30\% of its basic size at 90 BPM in local, local-server, and remote-server respectively. However, by virtue of accurate modeling and rigorous constraints applied to real-time interactive transformer tasks, the \sys is capable of stably identifying configurations that fulfill real-time requirements under complex and dynamic conditions.

\subsection{Key Metrics}
We list various metrics here to evaluate the performance of our system in terms of real-time performance and music quality.

\begin{table*}[tbp]
    \centering
        \caption{Experiment Results: Comparative Performance across Real-Time Settings at BPM = 120. Results are presented for three settings: (a) Local, (b) Local-server, and (c) Remote-server. Each row represents a configuration defined by the inference interval ($I$) and generation length ($GL$). Metrics are divided into Music Quality (JSD, FMD, CR, UR, $\text{NLL}_{\text{wavg}}$) and System Performance (ISR, Staleness, $\text{ISR}_w$). The Offline baseline serves as the upper bound for music quality. For generated samples, the best and second-best scores in each column are shown in boldface and underlined, respectively. $\downarrow$ indicates lower is better, and $\uparrow$ indicates higher is better.}
    \label{tab:quality-results}

\begin{subtable}{0.95\textwidth}
    \centering
    \caption{Local}
    \label{tab:local-results}
    
    \resizebox{\textwidth}{!}{%

        \begin{tabular}{l ccc ccc ccc}
            \toprule
            \multicolumn{1}{c}{} & \multicolumn{6}{c}{\textbf{Music Metrics}} & \multicolumn{3}{c}{\textbf{System Metrics}} \\ 
            \cmidrule(lr){2-7} \cmidrule(lr){8-10}
            $(I, GL)$ & JSD(Pitch) $\downarrow$ & JSD(Onset) $\downarrow$ & FMD $\downarrow$ & CR $\uparrow$ & UR $\downarrow$ & $\text{NLL}_{\text{wavg}}$ $\downarrow$ & ISR $\uparrow$ & Staleness $\downarrow$ & $\text{ISR}_w$ $\uparrow$ \\
            \midrule
            (1, 3) & \textbf{0.2941} & \textbf{0.1682} & 279.2 & \textbf{0.8207} & \textbf{0.0418} & \underline{1.498} & \textbf{0.9952} & \textbf{0.001} & \textbf{0.9951} \\
            (2, 5) & 0.3206 & 0.1904 & 269.5 & 0.7323 & 0.1012 & 1.690 & \underline{0.9781} & \underline{0.997} & \underline{0.9477} \\
            (2, 9) & 0.3354 & 0.3321 & 329.3 & 0.4693 & 0.4409 & 1.578 & 0.5795 & 1.944 & 0.5443 \\
            (4, 5) & 0.3831 & 0.2343 & 289.3 & 0.4758 & 0.3159 & 1.576 & 0.6855 & 1.230 & 0.6592 \\
            (4, 9) & 0.3075 & 0.1834 & \textbf{259.4} & 0.7247 & 0.1038 & 1.688 & 0.9511 & 2.419 & 0.8792 \\
            (4, 15) & \underline{0.3000} & \underline{0.1800} & 290.8 & \underline{0.7759} & \underline{0.0932} & 1.780 & 0.9495 & 3.783 & 0.8373 \\
            (7, 9) & 0.3618 & 0.3492 & 373.4 & 0.3413 & 0.5037 & \textbf{1.315} & 0.6160 & 2.455 & 0.5687 \\
            (7, 15) & 0.3234 & 0.1976 & \underline{262.5} & 0.6595 & 0.1411 & 1.806 & 0.8628 & 4.425 & 0.7436 \\
            Offline & 0.1960 & 0.1111 & 84.0 & 0.8257 & 0.0507 & 1.586 & - & - & - \\
            \bottomrule
        \end{tabular}}
    \end{subtable}
    
    \vspace{0.5em} 

    \begin{subtable}{0.95\textwidth}
    \centering
    \caption{Local-server}
    \label{tab:local-server-results}
    
    \resizebox{\textwidth}{!}{%

        \begin{tabular}{l ccc ccc ccc}
            \toprule
            \multicolumn{1}{c}{} & \multicolumn{6}{c}{\textbf{Music Metrics}} & \multicolumn{3}{c}{\textbf{System Metrics}} \\ 
\cmidrule(lr){2-7} \cmidrule(lr){8-10}
            $(I, GL)$ & JSD(Pitch) $\downarrow$ & JSD(Onset) $\downarrow$ & FMD $\downarrow$ & CR $\uparrow$ & UR $\downarrow$ & $\text{NLL}_{\text{wavg}}$ $\downarrow$ & ISR $\uparrow$ & Staleness $\downarrow$ & $\text{ISR}_w$ $\uparrow$ \\
            \midrule
            (1, 3) & 0.3456 & 0.4132 & 425.7 & 0.3292 & 0.5985 & 1.418 & 0.3813 & \textbf{0.544} & 0.3749 \\
            (2, 5) & 0.3172 & \textbf{0.1880} & \underline{263.4} & \textbf{0.7403} & \textbf{0.1082} & \underline{1.273} & \textbf{0.9536} & \underline{1.451} & \textbf{0.9104} \\
            (2, 9) & 0.4876 & 0.5396 & 733.1 & 0.0188 & 0.9733 & \textbf{1.079} & 0.0373 & 3.450 & 0.0333 \\
            (4, 5) & 0.4096 & 0.2680 & 325.3 & 0.4393 & 0.3633 & 1.485 & 0.5666 & 1.529 & 0.5395 \\
            (4, 9) & \textbf{0.3111} & \underline{0.2000} & \textbf{262.9} & \underline{0.6145} & \underline{0.1680} & 1.679 & \underline{0.7776} & 2.977 & \underline{0.7052} \\
            (4, 15) & 0.3571 & 0.4665 & 439.8 & 0.2218 & 0.7190 & 1.401 & 0.2827 & 5.281 & 0.2360 \\
            (7, 9) & 0.3798 & 0.3856 & 389.3 & 0.2562 & 0.6166 & 1.530 & 0.5131 & 2.805 & 0.4681 \\
            (7, 15) & \underline{0.3167} & 0.2169 & 354.7 & 0.4700 & 0.3516 & 1.979 & 0.6384 & 5.355 & 0.5316 \\
            Offline & 0.1960 & 0.1111 & 84.0 & 0.8257 & 0.0507 & 1.586 & - & - & - \\
            \bottomrule
        \end{tabular}}
    \end{subtable}

    \vspace{0.5em} 
    
    \begin{subtable}{0.95\textwidth}
    \centering
    \caption{Remote-server}
    \label{tab:remote-results}
    
    \resizebox{\textwidth}{!}{%

        \begin{tabular}{l ccc ccc ccc}
            \toprule
            \multicolumn{1}{c}{} & \multicolumn{6}{c}{\textbf{Music Metrics}} & \multicolumn{3}{c}{\textbf{System Metrics}} \\ 
\cmidrule(lr){2-7} \cmidrule(lr){8-10}
            $(I, GL)$ & JSD(Pitch) $\downarrow$ & JSD(Onset) $\downarrow$ & FMD $\downarrow$ & CR $\uparrow$ & UR $\downarrow$ & $\text{NLL}_{\text{wavg}}$ $\downarrow$ & ISR $\uparrow$ & Staleness $\downarrow$ & $\text{ISR}_w$ $\uparrow$ \\
            \midrule
            (1, 3) & 0.3741 & 0.4041 & 832.7 & 0.0167 & 0.9747 & \textbf{1.068} & 0.0115 & \textbf{1.000} & 0.0111 \\
            (2, 5) & 0.3093 & 0.2139 & 271.1 & 0.6568 & 0.1720 & 1.670 & 0.8679 & \underline{1.471} & 0.8280 \\
            (2, 9) & 0.3222 & 0.3978 & 356.4 & 0.3171 & 0.6020 & 1.510 & 0.4397 & 2.410 & 0.4066 \\
            (4, 5) & 0.4139 & 0.2876 & 326.2 & 0.3792 & 0.4410 & 1.480 & 0.5460 & 1.573 & 0.5191 \\
            (4, 9) & 0.3160 & \underline{0.1880} & \textbf{257.7} & \underline{0.7010} & \underline{0.1136} & \underline{1.306} & \underline{0.9376} & 2.518 & \underline{0.8638} \\
            (4, 15) & \textbf{0.3044} & \textbf{0.1639} & 285.4 & \textbf{0.7939} & \textbf{0.0572} & 1.351 & \textbf{0.9894} & 3.589 & \textbf{0.8785} \\
            (7, 9) & 0.3935 & 0.3805 & 391.1 & 0.2591 & 0.6103 & 1.527 & 0.5721 & 2.586 & 0.5259 \\
            (7, 15) & \underline{0.3078} & 0.2071 & \underline{262.0} & 0.6781 & 0.1408 & 1.764 & 0.8709 & 4.364 & 0.7522 \\
            Offline & 0.1960 & 0.1111 & 84.0 & 0.8257 & 0.0507 & 1.586 & - & - & - \\
            \bottomrule
        \end{tabular}}
    \end{subtable}

\end{table*}

\textbf{Real-time Performance} We evaluate the system performance by three metrics: Interaction Success Rate (ISR), Staleness and Weighted Interaction Success Rate ($\text{ISR}_w$).

ISR measures the proportion of time steps where \sys\, successfully find a piece of accompaniment to schedule and avoid event missing. First, we define an indicator function $I(t)$ for each time tick $t$: 
\begin{equation*}
    I(t) =
    \begin{dcases}
        1, & \text{if any notes can be scheduled at time $t$ } \\
        0, & \text{otherwise}
    \end{dcases}.
\end{equation*}
$I(t)=1$ means that the system is able to respond to the melody timely, no matter using backup notes or not. Then, the ISR can be defined as the average of this indicator function over the total sequence duration $N$:
$$
ISR = \frac{1}{N}\sum^{N-1}_{t = 0} I(t).
$$
A higher ISR indicates a more consistently available and responsive system, implying fewer interruptions or missing in the accompaniment generation and scheduling pipeline. \\
The Staleness metric quantifies the freshness of the scheduled note events. In a low-latency system, a scheduled note should ideally be one of the first tokens produced in the current generation step.
We define $S(t)$ as the Staleness Level at time $t$ when a note is scheduled:
\begin{align*}
    S(t) = &\text{ scheduled note's index in its generation sequence}.
\end{align*}
The overall Staleness is the average of $S(t)$ over all successful interaction time steps:
$$
Staleness = \frac{1}{|\{t: I(t)=1\}|}\sum_{t \in \{t:I(t) = 1\}} S(t).
$$
A lower Staleness means the system is quickly identifying and responding accompaniment request, resulting in fresher auditory feedback and better rhythmic alignment. Higher Staleness implies the system is slowly responding to the requests, forcing itself to play stale notes for each request.\\
The Weighted Interaction Success Rate ($\text{ISR}_w$) is a comprehensive metric that integrates both the frequency of successful interactions and response freshness. It is formally defined as:
$$
ISR_w = \frac{1}{N} \sum \left(\frac{C - S(t)}{C} \times I(t)\right).
$$
where the term $\frac{C-S(t)}{C}$ acts as a staleness penalty for a successful interaction $I(t)$. A higher  $\text{ISR}_w$ indicates the
system is not only good at responding timely,
but also delivering fresher responses. $\text{ISR}_w$ is crucial for comparing different scheduling strategies in a real-time environment. We set the constant $C=32$, corresponding to the maximum expected length of a generation sequence or frame.

\textbf{Music Quality} We evaluate the accompaniment generated by our models from two main aspects: the authenticity of the generated music and the harmonic alignment between the accompaniment and the conditioning melody. For the first aspect, we use four metrics: The Fr\'{e}chet Music Distance (FMD)~\cite{retkowski2025frechetmusicdistancemetric}, Negative Log-Likelihood (NLL), and Jensen-Shannon Divergence (JSD) for both Pitch (JSD(P)) and Onset (JSD(O)). FMD measures the statistical distance between distributions of features extracted from real and generated symbolic music, 
lower values indicate that the generated samples are closer in distribution to real performances. $\text{NLL}_{\text{wavg}}$ quantifies the internal fluency and statistical coherence of the accompaniment generated by our real-time system. 
Lower $\text{NLL}_{\text{wavg}}$ suggests that the generated music is statistically more consistent and fluent according to the model's learned distribution. Furthermore, we use the JSD to compare the statistical distributions of generated and real accompaniment: $\text{JSD}(\text{P})$ (Pitch) quantifies the fidelity of the note vocabulary used, and $\text{JSD}(\text{O})$ (Onset) evaluates the accuracy of the rhythmic and temporal patterns; lower values for both indicate better fidelity.  For the second aspect, the harmonicity metric evaluates how well the generated accompaniment supports the melody over time. For each melody note, we check all simultaneous accompaniment notes: if none overlap, that duration is counted as ``unsupported''. When overlap exists, we compute pitch-class intervals relative to the melody and compare them against a designated consonant set (unison, minor/major thirds, perfect fifth, and minor/major sixth). Durations with at least one consonant interval are labeled ``consonant'', while others are ``dissonant''. Normalizing these by the total melodic duration yields three distinct ratios: Consonant Rate (CR), Unsupported Rate (UR), and the Dissonant Rate. These ratios reflect how often the accompaniment harmonically supports, fails to reinforce, or conflicts with the melody, respectively. For the purpose of our study, we choose to present CR and UR in our results, as the Dissonant Rate can be trivially calculated from the other two ($1 - \text{CR} - \text{UR}$).

\subsection{Results}
\label{sec5:E}

Our evaluation focuses on three key areas: establishing the correctness of our system's performance metrics relative to generative coherence, analyzing the correlation between system success and music quality, and demonstrating the general feasibility of real-time accompaniment across various constraints.

\textbf{Correctness of The Proposed Model}
Through extensive experimentation, we conclusively validate the correctness of the solution space derived from our mathematical model, which predicts feasible real-time configurations. To demonstrate this accuracy, we utilize the Interaction Success Rate ($\text{ISR}$), which quantifies the system's reliability and successful delivery rate, as the core experimental measure of system viability. The experimental data provides robust confirmation for our theoretical predictions. For example (see Fig. \ref{fig:solution_space}), under the local setting and BPM = 120, multiple configurations that fall within our mathematically defined solution space—such as $(I=2, GL=5), \space (I=4, GL=9), \space(I=4, GL=15)$, and $(I=7, GL=15)$ —all achieve universally high $\text{ISR}$ values of 0.9781, 0.9511, 0.9495, and 0.8628 respectively. This strong consistency, where configurations predicted as feasible by our model exhibit high real-time success ($\text{ISR}$), provides direct and empirical evidence supporting the fundamental correctness of our mathematical solution space. This outcome confirms that our theoretical framework effectively and accurately identifies the optimal settings for real-time system deployment.

\textbf{Relations between System Performance and Music Quality} The experimental results establish a crucial synergistic relationship: successful system performance ($\text{ISR}_w$) acts as a strong, reliable proxy for high music quality. This is first evidenced by the validating relationship between $\text{ISR}_w$ and generative coherence ($\text{NLL}_{\text{wavg}}$). High $\text{ISR}_w$ correlates strongly with low $\text{NLL}_{\text{wavg}}$, confirming that reliable system delivery maintains the model's intrinsic statistical fluency. Conversely, the data reveals that $\text{NLL}_{\text{wavg}}$ is shown to be invalidated when $\text{ISR}_w$ is low. For example, in Table \ref{tab:local-server-results}, a scenario with the lowest $\text{NLL}_{\text{wavg}}$ ($\mathbf{1.079}$) registers a low $\text{ISR}_w$ ($\mathbf{0.0373}$). We believe the primary cause of this confusing phenomenon is that a system failure state yields sparse sequences containing a high proportion of ``empty'' tokens. When these sparse sequences are measured, the $\text{NLL}_{\text{wavg}}$ is artificially low, failing to reliably indicate the quality of the few generated notes. Therefore, the invalidation of $\text{NLL}_{\text{wavg}}$ when $\text{ISR}_w$ is low further underscores that system success is the prerequisite for generative metrics to accurately reflect quality. $\text{NLL}_{\text{wavg}}$ is a reliable indicator of generative coherence only above a high $\text{ISR}_w$ threshold. Beyond generative coherence, the data confirms a strong positive correlation between system success ($\text{ISR}_w$) and most objective musical quality metrics. As $\text{ISR}_w$ improves (increases), the music quality significantly improves across multiple domains: the JSD(Onset) (rhythmic timing) decreases, the Consonance Ratio (CR) increases, and the Undergeneration Rate (UR) decreases. This strong, multi-metric correlation validates $\text{ISR}_w$ as a robust proxy for real-time music performance, confirming that successful, low-latency delivery of notes inherently leads to superior musical results. For instance, the best local setting (Table~\ref{tab:local-results}, Row (1, 3)) achieves an $\text{ISR}_w$ of $\mathbf{0.9951}$ alongside high CR ($\mathbf{0.8207}$), excellent timing metrics, and an acceptable FMD, clearly demonstrating this synergy. The exception, JSD(Pitch), shows minimal correlation as it reflects the model's static harmonic vocabulary rather than the dynamic success or timing measured by $\text{ISR}_w$. 

\textbf{Feasibility in General} The experiments across the three tables (Local, Local-server, Remote-server) confirm the general feasibility of establishing various successful real-time accompaniment systems. While server and remote latencies significantly reduce the system's success rate, the system consistently provides a comprehensive solution space of feasible $(I, GL)$ configurations. For instance, even under challenging remote conditions (Table~\ref{tab:remote-results}), we identified solutions that maintain a functional $\text{ISR}_w$ of $\mathbf{0.8785}$, proving the robustness of our architecture against diverse network constraints. 

Crucially, the experimental results also clarify the optimal $(I, GL)$ selection strategy. First, maximizing interaction frequency requires the smallest possible Inference Interval ($I$). Second, a critical trade-off exists for Generation Length ($GL$). While a larger $GL$ safely buffers against network tail latency, it increases server inference latency and forces the model to predict too far ahead. This ignores recent user inputs, degrading freshness and coherence. Therefore, $GL$ should not be arbitrarily large; it should be kept just long enough to cover network jitter. This directly supports our core conclusion: keeping both $I$ and $GL$ as small as the system allows is important for balancing safety and real-time responsiveness.

For example, in the local-server setting (Table~\ref{tab:local-server-results}), the smaller-$I$ configuration $(2, 5)$ achieves the strongest overall system performance ($\text{ISR}=0.9536$, $\text{ISR}_w=0.9104$), illustrating the first principle in a concrete case. For $GL$, comparing $(4, 9)$ with $(4, 15)$ shows why larger is not always better: although both use the same $I$, increasing $GL$ sharply worsens freshness and coherence, with staleness rising from $2.977$ to $5.281$ and $\text{ISR}_w$ dropping from $0.7052$ to $0.2360$, while music quality also deteriorates (e.g., FMD from $262.9$ to $439.8$ and CR from $0.6145$ to $0.2218$).

\section{Discussion}
\label{Sec:discussion}

\subsection{Environmental Robustness and Adaptivity}
The RTT models fitted for the network-dependent environments (see section \ref{sec:5b}) represent a temporal ``snapshot'' of specific conditions during our benchmarking. Real-world network latency is highly influenced by external factors, such as time of day, concurrent traffic from other users, and hardware variability. Consequently, re-benchmarking these environments under different network conditions would likely yield a different set of model parameters. As we have seen in the previous visualization of the solution space (see Fig. \ref{fig:solution_space}), this means the precise boundaries of the solution space are reflective of these specific benchmark conditions and may not be universally robust.

At present, \sys adopts a profile-then-execute strategy: we first benchmark the deployment environment, estimate its latency profile, and then select fixed runtime parameters accordingly. This design is appropriate for the settings considered in this paper, where device placement and network conditions are relatively stable. For short-term latency fluctuations, the backup mechanism already provides robustness without requiring parameter reconfiguration. We agree, however, that more volatile environments may benefit from dynamic runtime adaptation, which we leave as an important direction for future work.

\subsection{Practical Relaxation of Theoretical Constraints}
Our current system model employs two strict constraints (see section \ref{sec:4c}) to delineate the feasible solution space. These constraints are inherently conservative as they are based on the 95th percentile of $RTT$. In our evaluation, any configuration pair $(I, GL)$ that violates these bounds is excluded from the ``valid'' space. However, in real-world musical interaction, these constraints may be ``soft'' ones.  A marginal violation of the Real-Time Safety Constraint (e.g., $I$ being slightly smaller than $RT_{tick}^{95}$) does not necessarily lead to a systemic collapse; rather, it increases the probability of occasional request staling. Given our backup mechanism and the perceptual resilience of human listeners to millisecond-level jitters, such rare delays might be virtually unnoticeable.

This suggests that the solution space is not a binary territory but a trade-off between reliability and responsiveness. A developer might intentionally select a ``near-boundary'' configuration to prioritize ultra-low latency (minimal $I$) over absolute continuity, accepting a non-zero but managed miss rate in exchange for a more responsive jamming experience. Future work could explore a probabilistic formulation of these constraints to better quantify this trade-off.

\subsection{
The Generation Gap in Real-Time Inference
} 

Despite the success of StreamMUSE, significant challenges remain for real-time deployment. The strict nature of the task imposes severe constraints on both context availability and model robustness. Unlike offline generation, real-time systems cannot rely on pre-injected structural prompts. This forces the model to ``blindly guess'' the accompaniment during cold starts. Furthermore, physical constraints inevitably pollute the streaming context. Unpredictable network delays force the client to play backup frames. This creates an inconsistency between the music the user hears and the history saved on the server. Additionally, human timing variations and hardware jitter often result in ``off-tick'' or incorrect note inputs. Our model struggles with these noisy inputs because it was trained only on perfect data via Maximum Likelihood Estimation (MLE), leading to the exposure bias discussed in Section \ref{model_architicture}. Consequently, the model cannot easily recover from distorted contexts. A single timing error or misplaced note can trigger a chain reaction that causes the music quality to deteriorate quickly. Meanwhile, the model also lacks stylistic flexibility. It cannot adapt on the fly if a user's playing style differs from the training data, which limits its usability in open-ended environments.

Addressing these limitations represents a key direction for future work. Promising approaches include reinforcement learning-based training to mitigate exposure bias~\cite{wu2025, scarlatos2025}, and inference-time robustness techniques such as scheduled sampling or token dropout to improve resilience under noisy streaming contexts.

\subsection{
Limitations in Evaluation
} 

Our study exposes the limitations of relying solely on existing objective metrics to assess real-time generation. Crucially, current music quality metrics are fundamentally designed for offline evaluation, assuming a complete and flawlessly generated sequence. As a result, they fail to adequately penalize artifacts unique to real-time streaming, such as the system failing to keep up with the performer or generating sparse, empty accompaniment. For instance, as discussed in Section \ref{sec5:E}, metrics like the weighted negative log-likelihood ($\text{NLL}_{\text{wavg}}$) can be highly misleading if evaluated in isolation. When physical perturbations force the system into low-reliability states, the model may degenerate into sparse patterns (essentially missing accompaniment events) that yield artificially low, falsely ``good'' NLL scores, despite a complete loss of actual musicality. Therefore, traditional offline metrics are insufficient on their own; they must be strictly conditioned on high system reliability (e.g., a high $\text{ISR}_w$ threshold) to meaningfully reflect actual generative coherence in a streaming context.

\section{Conclusion}
\label{Sec:conclusion}


This work presents a case study that establishes useful principles for building responsive, time-critical interactive transformer systems. By exploring the use case in accompaniment generation, systematically analyzing latency, synchronization, and inference behavior under real-time constraints, we provide a practical rule book for developers and researchers seeking to bridge the gap between algorithmic capability and interactive responsiveness. 
Our proposed \sys, to our best knowledge, is the first real-time accompaniment system that poses no musical constraints to user-input melodies. By focusing on real-time and cross-hardware adaptivity, our findings highlight key trade-offs between model complexity, system design, and perceptual thresholds, offering concrete guidance for future work on AI systems that must think, react, and create in real time.
\section*{Acknowledgment}
\label{Sec:acknowledgment}

The authors thank Lekai Qian and Jiadong Zhang for their valuable contributions during the post-submission deployment and optimization phase of the live demonstration system. In particular, their efforts in improving the deployment pipeline and tuning the execution environment helped strengthen the real-time responsiveness and stability of the system presented in this paper.

\bibliography{references}
\bibliographystyle{IEEEtran}

\clearpage



\end{document}